\documentclass[10pt]{wlscirep}
\usepackage[utf8]{inputenc}
\usepackage[T1]{fontenc}
\usepackage{placeins}
\usepackage{setspace}
\usepackage{lineno}
\usepackage[ruled]{algorithm2e}
\usepackage{soul}

\usepackage{graphicx} 
\usepackage{pdfpages}
\usepackage{tikz}
\usepackage{amsthm}
\usepackage{amsmath}
\usepackage{amssymb}
\usepackage{placeins}
\usepackage{xcolor}
\usepackage{svg}

\usepackage{caption}
\usepackage{bm}

\title{Uncertainty-triggered wake-up enables energy-efficient, error-resilient edge AI with memristor front ends}

\author[1]{Th\'eo Ballet} \author[2,3]{Aymen Romdhane} \author[2]{Bruno Lovison-Franco} \author[1,4]{Th\'eo Dupuis}
\author[1]{Adrien Renaudineau} \author[2]{Felipe Paiva Alencar} \author[1]{Mohammed Akib Iftakher} 
\author[1]{Cl\'ement Turck} \author[5]{Kamel-Eddine Harabi}
 \author[5]{Elisa Vianello}
\author[4]{Jean-Michel Portal} \author[2]{Pascal Benoit}  \author[2,*]{David Novo}  \author[1,*]{Damien~Querlioz} 
\affil[1]{Universit\'e Paris-Saclay, CNRS, Centre de Nanosciences et de Nanotechnologies,  Palaiseau, France}
\affil[2]{LIRMM, Universit\'e de Montpellier, CNRS, Montpellier, France.}
\affil[3]{Universit\'e Grenoble-Alpes, CEA, CNRS, Spintec,   Grenoble, France}
\affil[4]{Aix-Marseille Université, CNRS, Institut Matériaux Microélectronique Nanosciences de Provence, Marseille, France}
\affil[5]{Universit\'e Grenoble-Alpes, CEA, LETI,   Grenoble, France}
\affil[*]{david.novo@lirmm.fr, damien.querlioz@universite-paris-saclay.fr}

\begin{abstract}
Memristor computing offers a route to low-energy edge AI, but device variability, sensitivity to operating conditions, and system-integration challenges  can hinder deployment. Here we show that these limitations can be mitigated by using memristor AI not as the final decision maker but as the  ultra-low-power, always-on front end of a heterogeneous inference system. We implement this architecture by coupling a fabricated memristor Bayesian machine to a programmable CPU running a higher-power, higher-accuracy software neural network. The memristor front end acts as a probabilistic screener. When it predicts an abnormal event or produces an ambiguous or invalid output, a dedicated hardware wake-up path activates the CPU, which produces the final decision. We validate this architecture on a heartbeat-classification benchmark by interfacing the fabricated Bayesian machine with an FPGA-based wake-up platform and CPU back end. The resulting uncertainty-triggered wake-up system achieves high final classification accuracy under nominal operation and maintains this accuracy even when the memristor front end is degraded by voltage scaling or reduced programming margins, because unreliable outputs are converted into recoverable wake-up events instead of becoming silent errors. Post-layout  analysis of an ASIC implementation shows  that average energy is governed primarily by wake-up frequency, providing practical design rules for choosing front-end operating points. These results establish uncertainty-triggered wake-up as a  strategy for energy-efficient, error-resilient edge AI.
\end{abstract}

\begin{document}
\maketitle


\thispagestyle{empty}


\section*{Introduction}

Computing near or in memory with non-volatile nanodevices can drastically reduce the energy cost of artificial intelligence (AI) inference by alleviating the von Neumann data-movement bottleneck, which dominates many edge-AI workloads\cite{ielmini2018memory}. This promise has motivated demonstrations across a wide range of models and device classes. Memristor crossbars have been used for neural networks\cite{kim20214k,wan2022compute,hung2021four,huang2024memristor,li2020cmos,yousuf2025layer,jebali2024powering}, Bayesian neural networks\cite{lin2023uncertainty,bonnet2023bringing,lin2025deep}, Ising-type optimizers\cite{CaiNatElec2020,Jiang2022memHNN,jiang2023efficient,kim2024ising,shan2024one,iftakher2026intrinsic}, and Bayesian reasoning engines\cite{harabi2023memristor,turck2025logarithmic}. Related edge-AI accelerators have also been reported using phase-change memories\cite{ambrogio2023analog,le202364}, two-dimensional materials\cite{sebastian2022two}, ferroelectric memories\cite{kim2022cmos}, and magnetoresistive random-access memory (MRAM) \cite{jung2022crossbar}.

Yet the main challenge is no longer the demonstration of isolated compute kernels, but turning nanodevice AI into a useful edge system. At the device level, nanodevice front ends remain affected by variability, limited precision, and strong sensitivity to operating conditions. At the system level, communication overheads, control orchestration, and software/hardware mismatches can erase the gains of a specialized accelerator once it is embedded in a broader platform, an instance of Amdahl's law. In parallel, microcontroller-class processors have become increasingly efficient for on-device inference\cite{di2022kraken,busia2024tiny}, raising the bar for any alternative hardware. The relevant question is therefore not whether a nanodevice engine can replace programmable compute, but how the two can be combined so that each handles the part of the workload for which it is best suited.

Here we show that wake-up control provides such a combination. Rather than asking memristor AI to make every final decision, we use it as an always-on screening stage. 
A memristor-based Bayesian reasoning engine\cite{harabi2023memristor,turck2025logarithmic} monitors the input continuously at ultra-low power and wakes up a programmable back end only when it produces an abnormal prediction or an output too ambiguous to finalize locally.
The back end, implemented here as an open-source  RISC-V central processing unit (CPU)\cite{schiavone2017slow}, then executes a richer software model and produces the final decision. Importantly, the resulting architecture is heterogeneous in both hardware and inference style: the front end is a fixed-topology, mixed-signal Bayesian classifier whose score structure exposes ambiguity at negligible cost, whereas the back end is a programmable digital processor executing a neural-network model optimized for final-decision accuracy. This paradigm changes what is required from the memristor front end. 
Stand-alone software-equivalent accuracy is no longer the primary objective; instead, the front end must reject easy cases locally and reliably escalate the suspicious or uncertain ones. Wake-up therefore does not merely reduce average energy: it makes imperfect memristor AI useful by converting many front-end errors or failures into recoverable events handled by programmable compute.

Bayesian front ends are particularly attractive in this heterogeneous setting, because ambiguity can be inferred directly from their class-score structure. In an event-driven system, ambiguity is therefore not a limitation; it becomes a local trigger for more accurate computation. The same mechanism allows the front end to operate in more aggressive low-energy regimes, including reduced supply voltage or weaker memristor programming, as long as the resulting rare failures are preferentially converted into wake-ups instead of silent misclassifications. This robustness-aware operating mode is conceptually similar to Razor-style design, in which occasional errors are tolerated to save energy\cite{ernst2003razor}, but applied here at the level of AI inference and wake-up policy.

We validate this concept experimentally by interfacing a fabricated memristor Bayesian machine to a field-programmable gate array (FPGA) implementation of the remainder of the system, including the RISC-V CPU, memories, interconnect, and wake controller. We evaluate the platform on a heartbeat-classification task and deliberately stress the nanodevice front end by sweeping its supply voltage and memristor programming conditions. To assess system-level efficiency beyond the split prototype, we also design a 22-nm application-specific integrated circuit (ASIC) version of the digital platform and evaluate it with post-place-and-route power analysis. 
The results show that average energy is governed mainly by the always-on front-end cost and the wake-up frequency, which yields concrete design rules for choosing front-end operating points and wake criteria.

Wake-up sensing has previously been explored with CMOS-only receivers and lightweight classifiers \cite{rovere20182,izumi2015normally}. Beyond CMOS, several works have highlighted how emerging nanodevices could support always-on front ends. Ref.\cite{gupta2019low} proposed a trainable wake-up module based on a hybrid CMOS-MoS$_2$ memtransistor neuromorphic architecture, in which an always-on classifier mediates between the sensor and the main processing unit. Ref.\cite{kumar2019neuromorphic} formulated support-vector machines to better match memtransistor crossbar computing and outlined a two-stage pipeline where an SVM serves as a wake-up detector for a heavier CNN back end. Ref.\cite{jung2022crossbar} demonstrated a 64$\times$64 MRAM crossbar for in-memory MAC operations and illustrated an always-on vision pipeline in which low-power face detection produces a wake-up signal to enable more expensive face authentication. While these studies establish the relevance of nanodevice-based front ends for wake-up scenarios, they primarily focus on the wake-up module or the in-memory computing kernel itself. 
Here, by contrast, we experimentally demonstrate the complete wake-up path of a heterogeneous system. A fabricated memristor Bayesian front end performs always-on probabilistic screening, and a programmable CPU back end executes the higher-accuracy final classifier only when awakened.
We show experimentally that the wake-up mechanism can compensate for memristor imperfection, and an ASIC-grounded analysis shows that average energy is governed primarily by wake-up frequency. This work reframes the role of nanodevice AI at the edge: its value may lie less in replacing software inference than in providing trustworthy, ultra-low-power screening that allows programmable compute to remain asleep most of the time.

\begin{figure}[h]
    \centering
    \includegraphics[width=.9\linewidth]{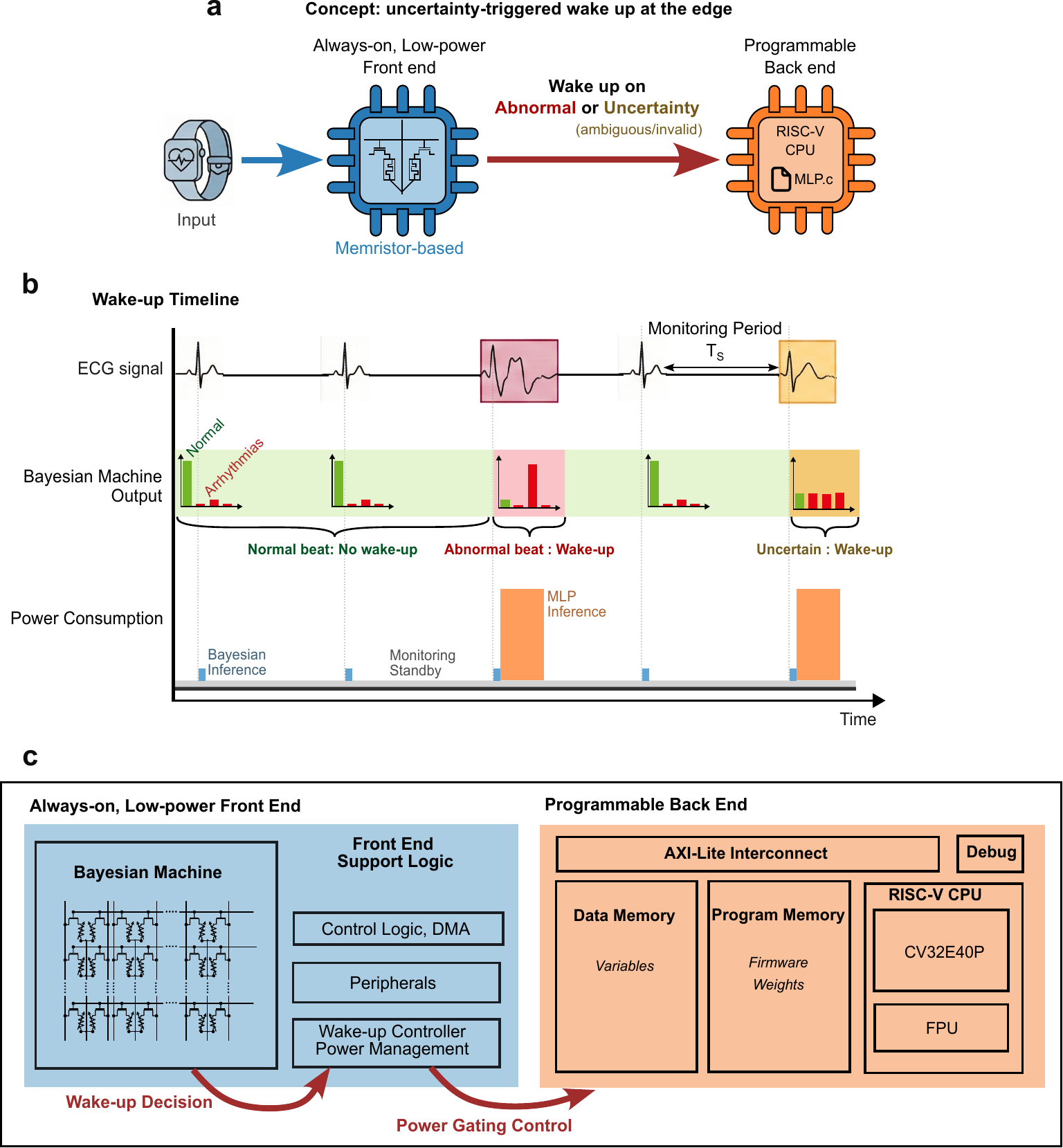}
\caption{\textbf{Heterogeneous uncertainty-triggered wake-up architecture.}
\textbf{a} Concept of the system. A memristor-based Bayesian front end performs always-on, ultra-low-power screening. Inputs confidently classified as normal are finalized locally, whereas abnormal predictions and ambiguous or invalid outputs trigger the programmable back end.
\textbf{b} Wake-up timeline. The front end evaluates successive inputs while the programmable back end remains asleep; wake-up temporarily activates the back end only for abnormal predictions or ambiguous/invalid outputs.
\textbf{c} System organization. The always-on front-end side contains the Bayesian machine and front-end support logic, including control/DMA, peripherals, and wake-up/power-management logic. The programmable back-end side contains the AXI-Lite interconnect, memories, debug logic, and the RISC-V CPU that runs the software multilayer perceptron (MLP). A dedicated hardware wake path links the two sides.
}
    \label{fig:cartoon}
\end{figure}

\FloatBarrier

\section*{Results}

\subsection*{Architecture and operating principle of the wake-up system}

Fig.~\ref{fig:cartoon} summarizes the heterogeneous wake-up architecture.
In the heartbeat benchmark studied in this work, each input corresponds to one heartbeat. 
A memristor-based Bayesian front end analyzes each input at ultra-low power (Figs.~\ref{fig:cartoon}a,b). Inputs classified as confidently normal are finalized locally, allowing the rest of the chip to remain asleep. Inputs predicted as abnormal (e.g., with arrhythmia), or yielding outputs too uncertain to finalize locally, trigger a wake-up request and are handed to a programmable back end for final classification. Such uncertain outputs may arise either because the input is genuinely difficult to classify or because front-end nonidealities produce an invalid output.

The wake-up concept itself does not require a Bayesian front end, but the Bayesian machine is particularly suitable because ambiguity can be inferred directly from its class-score structure with negligible additional hardware\cite{harabi2023memristor}. In this work, the programmable back end is implemented as a RISC-V CPU executing a software multilayer perceptron (MLP).
In the heartbeat implementation, the Bayesian machine and the MLP operate on two feature views of the same beat: a compact view for low-power screening by the front end, and a richer view for final classification by the back end (see below). The MLP is therefore not fed by the Bayesian-machine scores; it is invoked on the corresponding buffered input features when the front end requests wake-up.

The two stages are heterogeneous in both hardware and inference style. The always-on front end is a fixed-topology, mixed-signal Bayesian classifier with memristive storage and native uncertainty information, whereas the back end is a fully programmable digital processor that executes the MLP in software. 
The front end is therefore optimized for continuous ultra-low-power screening and selective escalation of difficult cases, while the back end is optimized for final accuracy and model flexibility (as the back-end model is defined in software, it can be updated over the lifetime of the device).

The organization is shown in Fig.~\ref{fig:cartoon}c. The always-on front-end side contains the Bayesian machine together with front-end support logic, including control and Direct Memory Access (DMA) functions, peripherals, and wake-up/power-management logic. The programmable back-end side contains the RISC-V CPU and its memories: a data memory and a program memory that stores both firmware (the software code executed by the CPU) and neural-network weights.  The back end also contains a lightweight on-chip communication fabric (an AXI-Lite bus), used by the CPU to access the memories, peripherals, and Bayesian-machine registers. A dedicated hardware wake path links the front-end and back-end sides.

As illustrated by the timeline of Fig.~\ref{fig:cartoon}b, the system spends most of its time in standby monitoring, where it consumes a low monitoring baseline power.
When the front end emits a wake-up signal, the power-management logic restores the programmable back end. The CPU then restarts from its standard firmware entry point. The first instructions determine whether this restart corresponds to a true system startup or to a wake-up request from the front end, and branch accordingly. 
In the latter case, the CPU immediately reads the front-end status and scores, retrieves the corresponding buffered MLP input features, executes the MLP inference, stores the final result, takes any necessary action, and returns the system to low-power monitoring.
Note that throughout the manuscript, we refer to each extracted beat segment as an input, and we call  $T_s$ the monitoring period between successive classifications.

\begin{figure}[h]
    \centering
   \includegraphics[width=\linewidth]{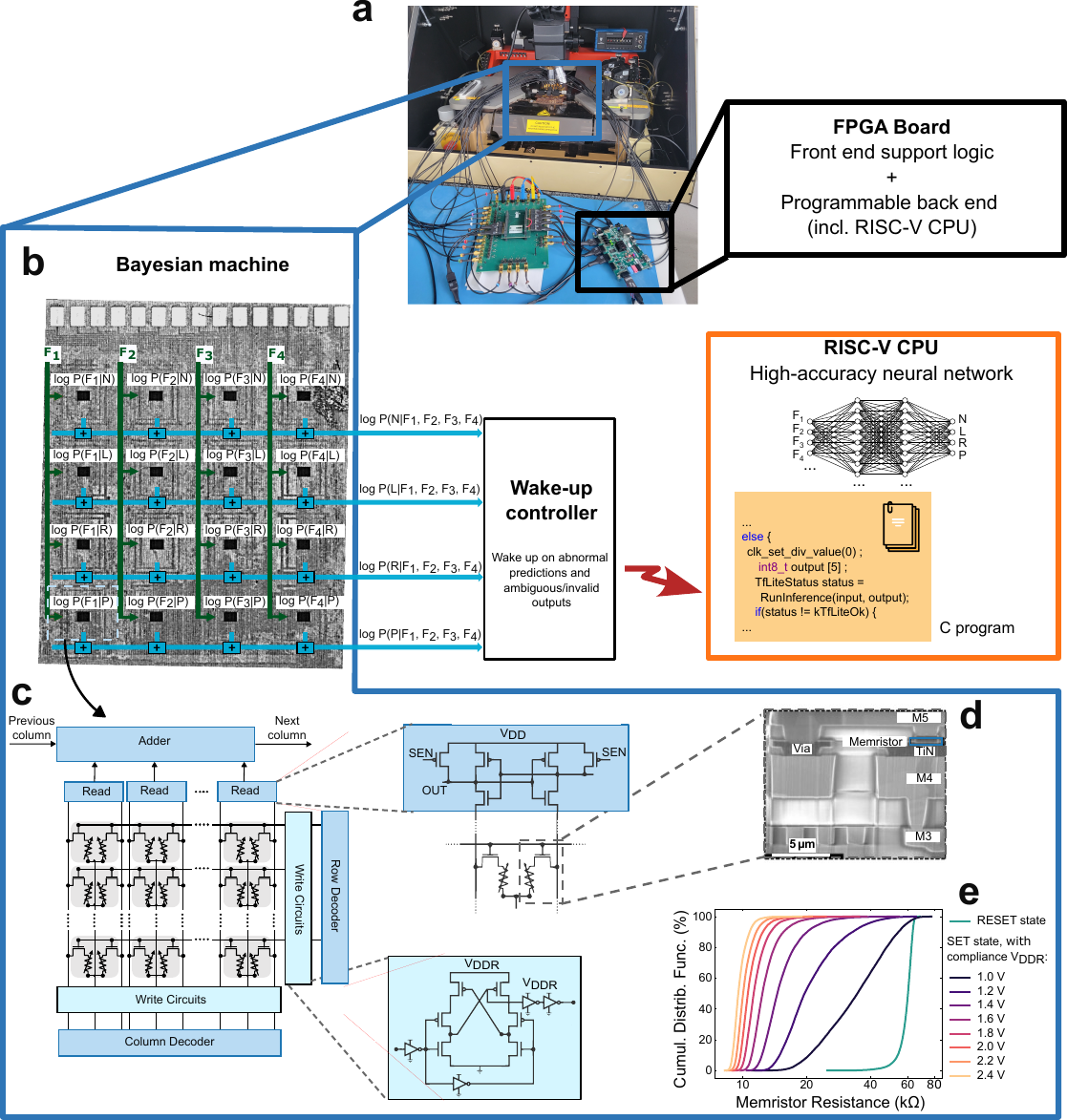}
   \caption{\textbf{Experimental platform and characterization of the memristor front end.}
\textbf{a} Photograph of the split prototype combining the fabricated logarithmic Bayesian machine on a probe station and the remainder of the platform implemented on FPGA.
\textbf{b} Die micrograph and functional overview of the Bayesian machine, showing the 16 memristor arrays used to accumulate class log-scores from the ECG-derived inputs and the logic that generates wake-up requests.
\textbf{c} Circuit details of the memory array and peripheral circuitry, including the 2T2R cell, precharge sense amplifier, and level shifter. The SET compliance is controlled through $V_\mathrm{DDR}$, whereas the Bayesian machine operates from $V_\mathrm{DD}$ during inference.
\textbf{d} Electron microscopy image of a memristor integrated above the CMOS stack.
\textbf{e} Cumulative resistance distributions after SET for different $V_\mathrm{DDR}$ values (measured on a separate programming-statistics test chip). Lower $V_\mathrm{DDR}$ narrows the programmed memory window and is therefore expected to degrade front-end robustness.
}
    \label{fig:machine}
\end{figure}

\FloatBarrier

We designed the complete system at the register-transfer level (RTL, see Methods) and used this design for both the experimental studies and the final-technology projections described below.

\subsection*{Uncertainty-triggered wake-up preserves accuracy across degraded front-end conditions}

To validate the complete wake-up path experimentally, we interfaced a fabricated logarithmic Bayesian machine to an FPGA implementation of the remainder of the system, including the RISC-V CPU, local memories, AXI-Lite interconnect, wake controller and rest of the front end support logic (Fig.~\ref{fig:machine}a,b). The fabricated front-end Bayesian machine is therefore exercised with its real device-level variability and analog non-idealities, while the digital wake-up path, reset-based wake handling, and back-end inference are evaluated cycle-accurately using the same RTL as in the ASIC study.
This split prototype lets us measure not only stand-alone front-end behavior, but the full system response to each beat in the test stream: local classification, wake-up generation, CPU service, and final decision.

The fabricated front end is a logarithmic Bayesian machine derived from ref.~\cite{turck2025logarithmic} (which is a logarithmic deterministic re-implementation of the stochastic Bayesian machine of ref.\cite{harabi2023memristor}). For the present four-class task, it uses 16 memristor arrays, corresponding to the $4\times4$ class-feature combinations of the selected ECG descriptors. During inference, each of the four quantized input features addresses one stored 8-bit log-likelihood word per class, and these contributions are accumulated to produce four class scores (see Methods). This organization is well suited to wake-up operation because the machine outputs not only a winning class, but the full score pattern from which ambiguity can be detected with negligible extra hardware.

The mixed-signal core combines differential two-transistor/two-resistor (2T2R) memristive storage, precharge sense amplifiers, and level shifters between the low-voltage inference domain and the higher-voltage programming domain (Fig.~\ref{fig:machine}c). In the operating-condition sweeps below, we vary one inference-time knob and one programming-time knob: the front-end supply voltage $V_\mathrm{DD}$ during inference, and the SET-compliance voltage $V_\mathrm{DDR}$ used during the preceding memristor-programming step. Although $V_\mathrm{DDR}$ is a programming-time rather than a runtime knob, it matters because it sets the conductance separation that the front end must later read during every inference. Lowering $V_\mathrm{DD}$ reduces dynamic energy roughly quadratically but degrades analog read margin and speed. Lowering the programming-time $V_\mathrm{DDR}$ narrows the programmed conductance window (Fig.~\ref{fig:machine}e), providing a controlled way to test how weaker programming margins propagate to wake-up behavior and final system accuracy. Throughout the main text, $V_\mathrm{DDR}$ therefore denotes the earlier programming condition, not a swept inference-time supply.

\begin{figure}[h]
    \centering
   \includegraphics[width=.9\linewidth]{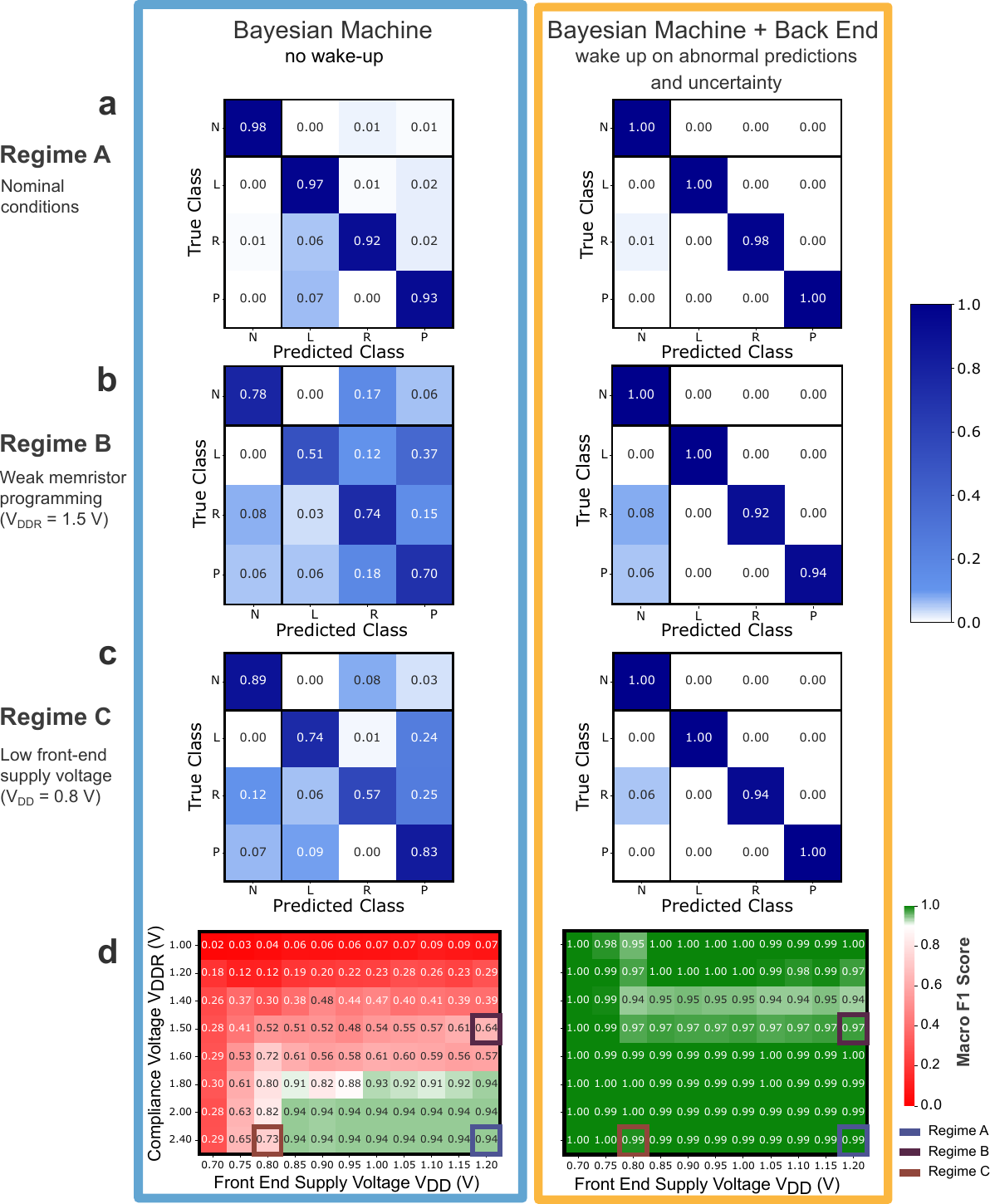}
\caption{\textbf{Uncertainty-triggered wake-up restores system-level accuracy across degraded front-end conditions.}
\textbf{a-c} Confusion matrices for three representative operating conditions. $V_\mathrm{DD}$ denotes the front-end supply voltage, and $V_\mathrm{DDR}$ the SET-compliance voltage used during the preceding memristor-programming step. Regime A, nominal operation ($V_\mathrm{DD}=1.2$~V, $V_\mathrm{DDR}=2.4$~V); regime B, weaker memristor programming ($V_\mathrm{DD}=1.2$~V, $V_\mathrm{DDR}=1.5$~V); and regime C, reduced front-end supply ($V_\mathrm{DD}=0.8$~V, $V_\mathrm{DDR}=2.4$~V).
For each regime, the left matrix shows the memristor Bayesian front end operating alone, whereas the right matrix shows the heterogeneous two-stage system, in which beats flagged as abnormal, ambiguous, or invalid are reclassified by the programmable MLP back end. N, normal; L, left bundle branch block; R, right bundle branch block; P, paced.
\textbf{d} Macro-F1 score across the three abnormal classes for all measured combinations of $V_\mathrm{DD}$ and $V_\mathrm{DDR}$, for front-end-only operation (left) and the full heterogeneous wake-up system (right).}
        \label{fig:experiments}
\end{figure}

\FloatBarrier

As a controlled proof-of-concept benchmark, we consider a four-class heartbeat-classification task (normal beats and three abnormal beat classes) derived from the public Massachusetts Institute of Technology-Beth Israel Hospital (MIT-BIH) Arrhythmia Database. To compare hardware operating conditions on exactly the same inputs, 
we stream prerecorded beats from the test set through the memristor Bayesian machine, 
while the RISC-V back end remains inactive unless explicitly awakened. The train/test split is obtained by random beat-level sampling (see Methods), so this task should be interpreted as a controlled benchmark for comparing operating conditions and wake-up policies, not as a patient-independent clinical evaluation. 
In this proof-of-concept, beat detection, segmentation, and FFT-based feature extraction are performed off-chip; the implemented wake-up platform therefore starts from feature vectors. The Bayesian machine receives four quantized FFT features for screening, whereas the MLP receives 32 quantized FFT features from the same beat when the back end is awakened (see Methods). The reported hardware and energy results therefore quantify the wake-up classification subsystem, not the full sensing chain.

Fig.~\ref{fig:experiments} shows the key classification result. Under nominal conditions (regime A), the front end already separates the four heartbeat classes reasonably well: the diagonal entries of the confusion matrix are 0.98, 0.97, 0.92, and 0.93 for normal beats (N), left bundle branch block beats (L), right bundle branch block beats (R), and paced beats (P), respectively. Most residual errors correspond to confusions, not abnormal beats classified as normal. For wake-up screening, this distinction matters: confusion among abnormal classes is usually benign because it still triggers escalation, whereas abnormal beats being finalized as normal are more dangerous because they can suppress wake-up altogether.

To compare classification quality across the full $V_\mathrm{DD}$, $V_\mathrm{DDR}$ sweep, overall accuracy is not the most informative metric. In realistic deployment streams normal beats dominate, so overall accuracy can remain high even when some abnormal beats are missed. We therefore report the macro-F1 over the three abnormal classes. For each class $c\in\{L,R,P\}$, we compute the per-class F1 score
\begin{equation}
    F1_c=\frac{2\,\mathrm{Precision}_c\,\mathrm{Recall}_c}{\mathrm{Precision}_c+\mathrm{Recall}_c},
\end{equation}
and we define
\begin{equation}
    \mathrm{Macro}\text{-}F1=\frac{F1_L+F1_R+F1_P}{3}.
\end{equation}
This classic metric gives equal weight to the three abnormal classes and is high only when both false positives and false negatives are well controlled for each of them. The abnormal-class macro-F1 of the front end under nominal conditions is 0.94. 

In this work, the wake criterion is deliberately hardware-light. Any beat whose top-scoring class is abnormal already triggers wake-up, so the only additional cases that matter are outputs that would otherwise be finalized locally as normal. We therefore add an ambiguity test: wake up when a normal-beat decision ties with at least one abnormal class score. Importantly, the wake-up mechanism itself does not require the front end to resolve which abnormal class is present: a binary normal/abnormal detector could already drive escalation. The advantage of retaining multi-class front-end scores is that it enables this tie-based ambiguity criterion at essentially no additional hardware cost. We also wake up on invalid output states, detected here when one or more decoded class probabilities are exactly zero. Because the smoothed Bayesian model contains no exact zero probabilities, such states are treated as hardware or numerical failures, not meaningful classifier outputs. 
 In the rest of the manuscript, we use ``uncertainty-triggered wake-up'' to refer to this complete hardware policy: wake-up on abnormal predictions, ambiguous normal predictions, or invalid outputs.

Once this policy is enabled, regime A becomes nearly ideal at the system level, with diagonal terms of 1.00, 1.00, 0.98, and 1.00. The abnormal-class macro-F1 rises from 0.94 to 0.99. In this situation 99.8\% of abnormal beats lead to a wake-up event (98.3\% due to an explicit abnormal prediction and 1.50\% due to an ambiguous or invalid output). Only 1.88\% of normal beats lead to a wake-up, all of them misrecognized as abnormal by the Bayesian machine and none due to an ambiguous or invalid output.

We next examine two degraded operating regimes in which front-end accuracy drops sharply: regime B uses a lower programming-time $V_\mathrm{DDR}$, which narrows the stored conductance window, and regime C uses a lower inference-time $V_\mathrm{DD}$, which reduces the read margin of the precharge sense amplifier during inference.

In regime B, the front-end-only abnormal-class macro-F1 falls to 0.64, with strong confusion between paced and bundle-branch-block beats. In regime C, it remains only 0.73, with missed R beats and substantial leakage of L and paced beats into competing classes. Once uncertainty-triggered wake-up is enabled, however, the final system recovers to 0.98 and 0.99 macro-F1, respectively. The corresponding wake-up burden is very different: the wake-up rate on normal beats rises to 22.5\% in regime B and 77.3\% in regime C; in regime C, this increase is driven almost entirely by ambiguous or invalid outputs. Regime C should therefore be interpreted as a deliberately over-scaled stress condition, not as an attractive deployment operating point. It shows that uncertainty-triggered wake-up can preserve final classification even for a severely degraded front end, but at the price of waking the CPU too often for good energy efficiency (see below).  The detailed wake-up fractions for representative regimes A-C, including the split between explicit abnormal predictions and ambiguity/invalid-output triggers, are summarized in Supplementary Table~1.

The key point is therefore not that the memristor front end remains accurate in isolation, but that it remains informative enough to reject easy cases locally while escalating difficult or unreliable ones.
Across the full measured $V_\mathrm{DD}$-$V_\mathrm{DDR}$ sweep, the abnormal-class macro-F1 remains between 0.94 and 1.00 with uncertainty-triggered wake-up, whereas front-end-only operation spans 0.02 to 0.94.

Overall, the the wake-up mechanism decouples correctness from strict analog fidelity, enabling aggressive $V_\mathrm{DD}$ scaling and weak programming conditions without sacrificing detection reliability. We next quantify the energy implications of these policies and identify the operating points that minimize mean energy per input for representative monitoring cadences.

\FloatBarrier

\begin{figure}[h]
    \centering
   \includegraphics[width=\linewidth]{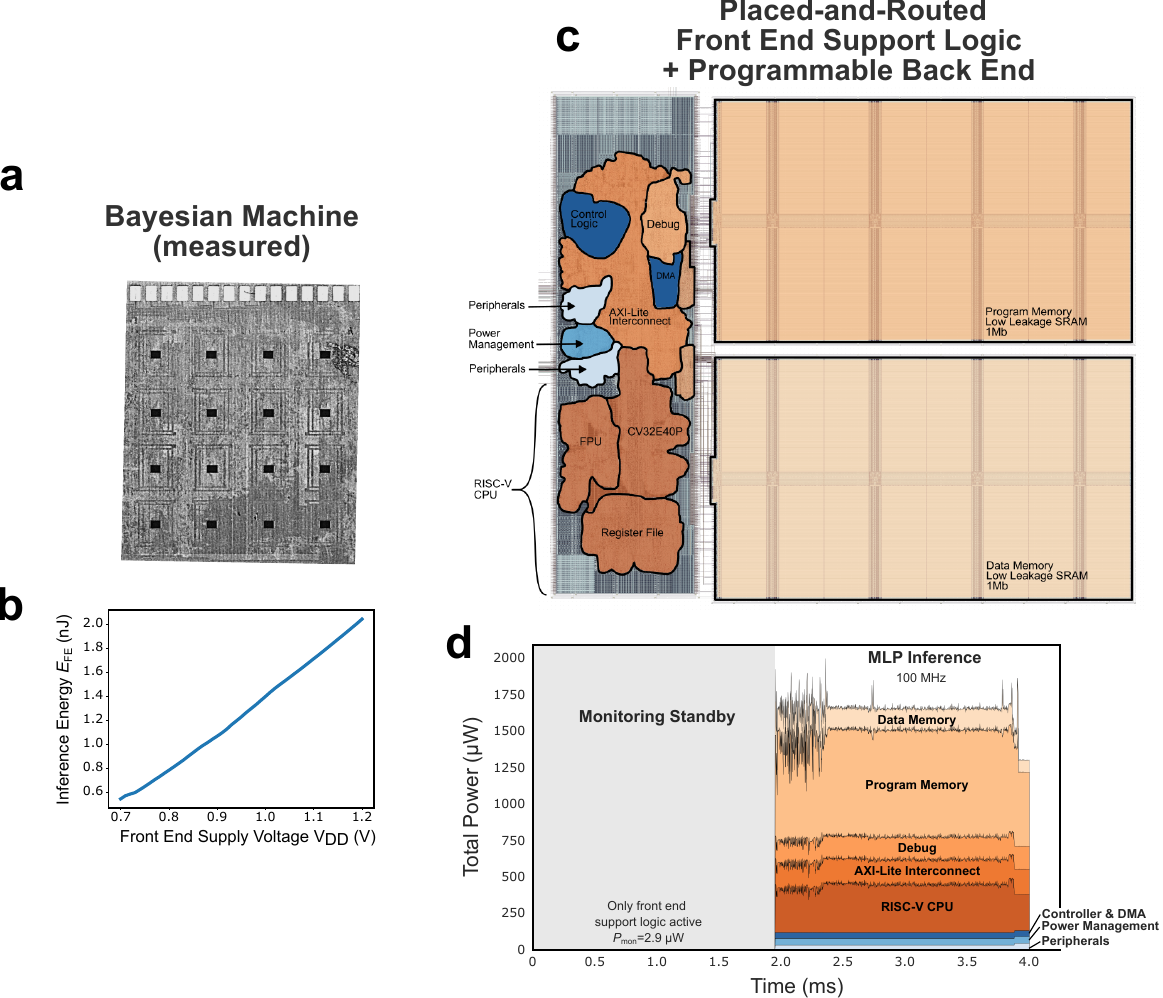}
      \caption{\textbf{Measured front-end characteristics and simulated post-layout digital platform used for system-level energy evaluation.}
\textbf{a} View of the fabricated Bayesian machine used to obtain measured front-end energy.
\textbf{b} Measured inference energy of the Bayesian machine versus front-end supply $V_\mathrm{DD}$.
\textbf{c} Post-place-and-route view of the 22-nm digital ASIC implementing the front-end support logic and programmable back end.
\textbf{d} Simulated time-resolved power breakdown during monitoring standby and a representative wake-up service event, including MLP inference.}
    \label{fig:asic}
\end{figure}

\FloatBarrier

\subsection*{Wake-up probability and monitoring cadence determine the energy optimum}

To estimate system-level energy beyond the split prototype, we combine measured  behavior of the Bayesian machine with a post-layout implementation of the digital back end and of the front end support logic. This analysis is intrinsically heterogeneous: the memristor Bayesian front end is characterized on the fabricated 130-nm CMOS+memristor hardware (Fig.~\ref{fig:asic}a,b) used in the previous section, whereas the front-end support logic and programmable back end are implemented from the same RTL as in the split prototype, in 22-nm FD-SOI, and analyzed after place and route (Fig.~\ref{fig:asic}c,d). The resulting numbers should therefore be interpreted as a system-level projection anchored to measured front-end hardware, not as the exact energy of a monolithic single-node chip.

This partition is also technologically natural. The front end is dominated by mixed-signal sensing, memristor integration, and relatively small non-volatile state storage, so its behavior is governed mainly by device physics and analog margins rather than by the most aggressive CMOS node. By contrast, the programmable back end benefits directly from digital scaling through denser memories, lower logic energy, and lower leakage in power-gated standby. A heterogeneously integrated implementation (e.g., using chiplets) is therefore a plausible deployment path.

Post-layout power is obtained by back-annotated gate-level simulation of the 22-nm implementation running the same firmware as in the experiments. The simulations include idle monitoring, restoration of the clock- and power-gated digital domains after wake-up, reset-based wake handling, and MLP inference. Switching activity is captured from delay-annotated simulations and combined with extracted parasitics for power analysis (see Methods). Fig.~\ref{fig:asic}b shows the measured inference energy of the Bayesian machine as a function of $V_\mathrm{DD}$, Fig.~\ref{fig:asic}c shows the placed-and-routed digital system on chip (SoC), and Fig.~\ref{fig:asic}d shows the time-resolved power breakdown during a representative wake-up service event. Together, these panels highlight the logic of event-driven operation: a very low-cost always-on front end (2.0~nJ/inference at nominal $V_{DD}$) is paired with a more accurate but much more expensive (3.2~$\mu$J per wake-up service episode, including restoration and MLP inference) programmable service path that is activated only when necessary.

Within a wake-up service event, the dominant contribution comes from the memory subsystem together with the AXI-Lite interconnect  (mostly used for memory access), which account for about 61.7\% of the total energy, consistent with the well-documented dominance of memory traffic during neural-network inference on programmable hardware. Most of this cost is associated with the neural-network weights, which are stored in the same program memory as the firmware. By comparison, the RISC-V CPU itself contributes only about 19.4\% of the energy. The joint test action group (JTAG) debug interface also contributes significantly in our prototype-oriented implementation, reaching about 10.7\%. This JTAG interface is not intrinsic to the proposed architecture and could be reduced or eliminated in a deployment-oriented ASIC.

\begin{figure}[h]
    \centering
   \includegraphics[width=0.9\linewidth]{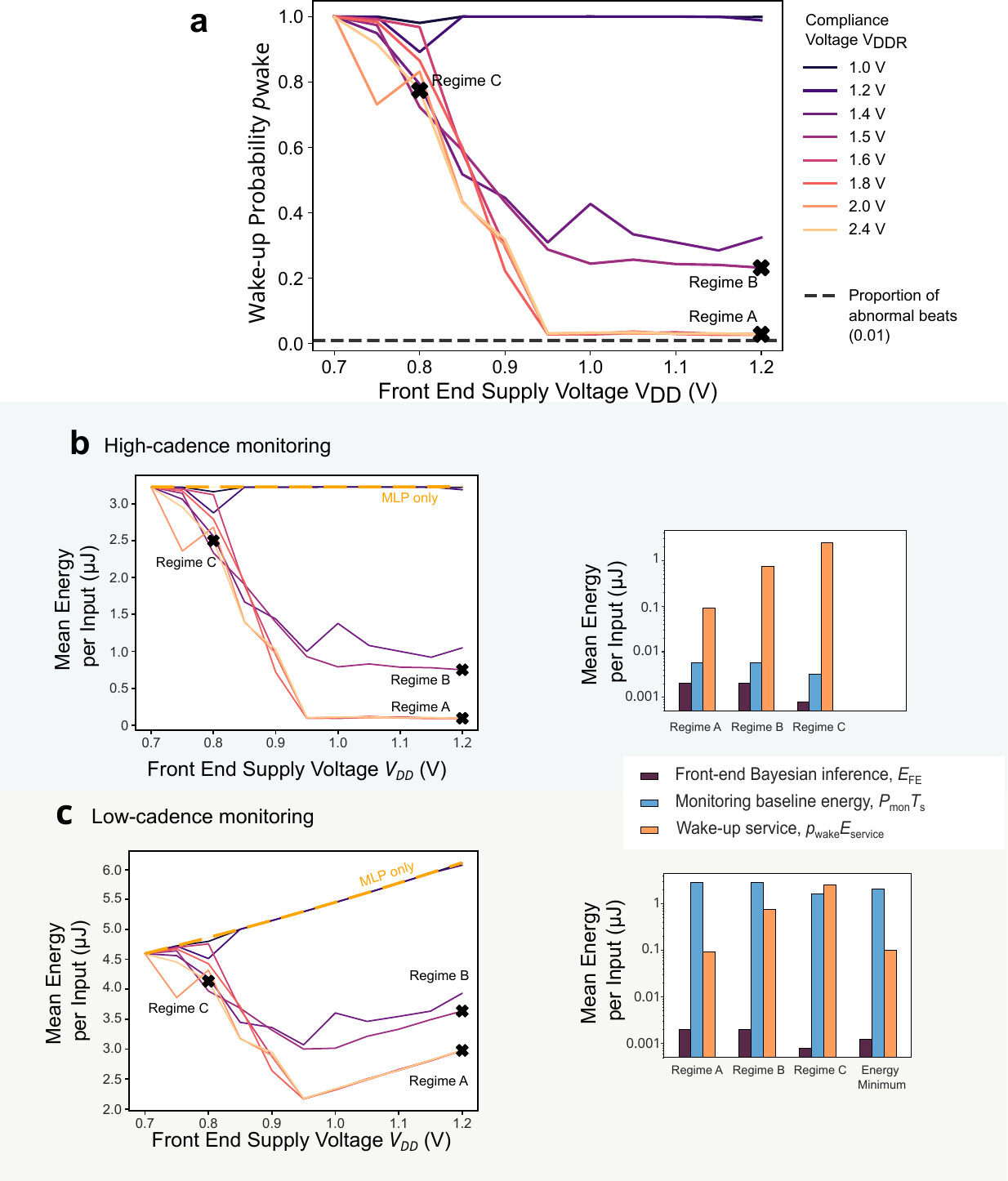}
\caption{\textbf{Wake-up probability and monitoring cadence govern mean energy per input.}
\textbf{a} Per-input wake-up probability $p_\mathrm{wake}$ versus front-end supply $V_\mathrm{DD}$ for the tested memristor programming conditions $V_\mathrm{DDR}$. The dashed line at $\pi=0.01$ marks the ideal floor for perfect screening (wake-up on all abnormal inputs and on no normal inputs). Crosses indicate operating points A-C from Fig.~\ref{fig:experiments}.
\textbf{b} Mean energy per input for a short monitoring period ($T_s=2$~ms). Left, energy versus $V_\mathrm{DD}$ for the tested $V_\mathrm{DDR}$ values, together with the baseline corresponding to running the MLP on every input. Right, decomposition of selected operating points into front-end inference, monitoring, and wake-up service (including restoration and MLP inference). 
\textbf{c} Mean energy per input for a long monitoring period ($T_s=1$~s). Left, corresponding energy sweep and MLP-only baseline. Right, decomposition of selected operating points. Here the monitoring term becomes comparable to the wake-up-service term, producing an interior optimum. All curves assume an abnormal-input probability $\pi=0.01$ and combine measured front-end inference energy with post-layout estimates for the 22-nm digital support logic and programmable back end.}
\label{fig:energy}
\end{figure}

\FloatBarrier

During the interval between two front-end classifications, the programmable back end is power- and clock-gated and the memristive Bayesian machine consumes negligible power, as memristors are non-volatile.
The remaining power is the monitoring baseline $P_\mathrm{mon}$, i.e. the power of the always-on front-end support domain that keeps the wake path, configuration/status registers, control logic, and peripheral interface available. In the as-implemented RTL and layout, $P_\mathrm{mon}=2.9\mu W$ at nominal $V_\mathrm{DD}$.  Approximately 55\% of $P_\mathrm{mon}$ is static and 45\% is dynamic switching. The dynamic component is not a fundamental lower bound: it includes residual clocked activity in the always-on support logic and could be reduced by more aggressive clock gating, a lower monitoring clock, or finer-grained power gating. The static component could be reduced by reverse back-biasing in FD-SOI. We retain the as-implemented value in the main analysis to quantify how a finite always-on monitoring baseline shifts the system-level energy optimum.

All confusion matrices and macro-F1 values reported in Fig.~\ref{fig:experiments} were computed on the balanced test set described in Methods.  For the energy analysis, we therefore reweight the measured class-conditional wake statistics to this deployment prior. Denoting by $p_\mathrm{wake|abn}$ and $p_\mathrm{wake|N}$ the wake probabilities on abnormal and normal inputs, respectively, the overall per-input wake-up probability is
\begin{equation}
p_\mathrm{wake}=\pi\,p_\mathrm{wake|abn}+(1-\pi)\,p_\mathrm{wake|N}.
\end{equation}
Because normal inputs dominate in this regime, the excess of $p_\mathrm{wake}$ above the ideal floor $\pi$ is driven mainly by unnecessary wake-ups on normal inputs. 

The mean energy per input can be written as
\begin{equation}
E_\mathrm{avg}=E_\mathrm{FE}+ E_\mathrm{mon} + p_\mathrm{wake}\,E_\mathrm{service}.
\label{eq:energie}
\end{equation}
Here, $E_\mathrm{FE}$ is the measured energy of one front-end inference, $E_\mathrm{service}$ is the energy of one wake-up and back-end service episode extracted from the 22-nm post-layout simulations, and $E_\mathrm{mon}$ is the contribution of the always-on digital support logic. The latter is negligible during active front-end and back-end inference and is dominated by the monitoring interval between successive inputs, which can be approximated by
\begin{equation}
E_\mathrm{mon} = P_\mathrm{mon}T_s,
\label{eq:monitorcost}
\end{equation}
where $T_s$ is the delay between successive inputs (which we call monitoring period). In the heartbeat benchmark, the physiologically relevant case is approximately one classification per beat. However, because the proposed wake-up architecture targets sparse-event edge sensing more broadly, we treat $T_s$ here as a sweepable parameter.

This energy decomposition makes the central trade-off explicit. Lowering $V_\mathrm{DD}$ is beneficial only insofar as it reduces 
the front-end energy  ($E_\mathrm{FE}+ E_\mathrm{mon}$) without inflating $p_\mathrm{wake}$; once voltage scaling or weaker programming causes more false or uncertainty-triggered wake-ups, the service term  $p_\mathrm{wake}\,E_\mathrm{service}$ dominates and total energy rises. Because one wake-up service episode costs 3.2~$\mu$J whereas one front-end inference costs only 2.0~nJ at nominal $V_\mathrm{DD}$, each additional 1 percentage point of $p_\mathrm{wake}$ adds about 32~nJ to the mean energy per input.

Fig.~\ref{fig:energy}a shows the wake-up probability  $p_\mathrm{wake}$ across the full $V_\mathrm{DD}$-$V_\mathrm{DDR}$ sweep. The nominal operating point A lies close to the ideal floor, with $p_\mathrm{wake}\approx 0.03$. By contrast, the weaker-programming operating point B reaches $p_\mathrm{wake}\approx 0.23$, and the over-scaled operating point C reaches $p_\mathrm{wake}\approx 0.78$.

Fig.~\ref{fig:energy}b looks at the mean energy per input when a short monitoring period $T_s$ is used between inputs (2~ms). In this case, the monitoring contribution  $P_\mathrm{mon}T_s$ is negligible, and the mean energy per input closely follows $p_\mathrm{wake}$. Operating point A therefore gives the lowest energy among the representative regimes (100~nJ/input), with a reduction by a factor 34 with regards to operating the MLP on each input.
Regime B is substantially worse because of its higher wake-up probability (757~nJ/input). Regime C (2.5~$\mu$J/input), as expected, loses most of the benefit of event-driven operation because excessive wake-ups make the service term dominant. 

 Fig.~\ref{fig:energy}c shows the corresponding behavior for a long monitoring period.  Here, $P_\mathrm{mon}T_s$ becomes significant, so the energy floor is no longer set by wake-up-service energy alone. Moderate reduction of $V_\mathrm{DD}$ can lower the front-end inference and monitoring energy costs without yet triggering many extra wake-ups, producing an interior optimum.  This energy minimum corresponds to an energy saving by a factor 2.4 with regards to operating the MLP on each input, assuming the monitoring is done at the same voltage. Pushing $V_\mathrm{DD}$ lower eventually increases $p_\mathrm{wake}$ enough that the wake-up-service term dominates again.

These results show a qualitative trend. In deployment-oriented systems, the monitoring power $P_\mathrm{mon}$ may vary extensively.
A deployment optimized for very sparse sensing could reduce $P_\mathrm{mon}$ using deeper sleep, lower monitoring frequency, state-retentive power gating, or back-biasing, allowing recovering energy savings similar to Fig.~\ref{fig:energy}b. Conversely, applications requiring continuous sensor biasing, interface activity, or safety monitoring may have a comparable or larger baseline. The relevant design parameter is therefore $P_\mathrm{mon}T_s$, not $P_\mathrm{mon}$ alone.

For completeness, Supplementary Fig.~1 extends this analysis to a broader sweep of $T_s$.

The design rule is therefore simple. First, choose programming conditions and wake-up criteria that suppress unnecessary wake-ups, especially on normal inputs. Second, lower $V_\mathrm{DD}$ only until the induced increase in $p_\mathrm{wake}$ cancels the savings in front-end inference energy. In other words, the right optimization target for the memristor front end is not maximal stand-alone accuracy, but minimal mean energy per input at fixed task-level reliability.

\FloatBarrier

\section*{Discussion}

This work demonstrates a heterogeneous wake-up system for edge AI that couples an always-on probabilistic memristor front end to a programmable neural-network back end. A memristor-based Bayesian machine monitors continuously at ultra-low power and issues a hardware wake signal either on abnormal predictions or on ambiguous or invalid outputs; an on-chip RISC-V CPU then executes a richer model only when needed. Experiments on heartbeat classification, together with an ASIC gate-level power analysis, show that average energy is governed primarily by wake-up frequency rather than by baseline front-end power alone. Uncertainty-triggered wake-ups decouple correctness from strict analog fidelity: 
we can lower $V_\mathrm{DD}$ and use weaker programming conditions while maintaining detection reliability because the back end arbitrates the rare ambiguous cases. 
The resulting system delivers large energy savings relative to continuous software execution and provides clear design rules for selecting wake criteria and operating points (Figs.~\ref{fig:experiments}-\ref{fig:energy}).

Wake-up receivers and front-end classifiers implemented purely in CMOS typically perform simple detection or shallow classification (e.g., energy/threshold detectors, SVMs), handing off to a microcontroller for full processing. 
Such designs minimize idle power but rarely expose directly usable ambiguity information. Our platform differs in two ways: (i) the front end is a nanodevice classifier whose class-score pattern supports a low-overhead ambiguity trigger; and (ii) we validate the complete path, from device to CPU, experimentally and with post-layout power analysis, turning component-level gains into system-level design guidance.

In this prototype, the front end is still modest (a 16-array, log-domain Bayesian machine), yet nanodevices are well suited to implement more advanced probabilistic models. Their stochastic and nonlinear characteristics map naturally onto Bayesian computations and can yield hardware that estimates uncertainty as a first-class output\cite{lin2023uncertainty,bonnet2023bringing,lin2025deep,querlioz2025bayesian}. This property is particularly valuable for event-driven systems: uncertainty becomes a reliable, local criterion for waking the programmable back end, tightening the energy–reliability loop without complex firmware. 
As front ends scale, e.g., to deeper Bayesian networks, structured priors, or multimodal fusion, they may reduce spurious wake-ups at a fixed detection target, further improving system energy.

Our results suggest a simple hierarchy. First, minimize unnecessary wake-ups; second, tune front-end ($V_\mathrm{DD}$) and programming to the knee where reductions in baseline power do not inflate the wake rate. Because the back end is fully programmable and updatable, application evolution does not require redesigning the nanodevice front end; only its operating thresholds or parameters need retuning. These rules extend beyond memristors to other non-volatile technologies (PCM, ferroelectrics, MRAM) and to alternative programmable back ends (e.g., MCU+DSP or NPU).

Another advantage of the proposed approach is that it can relax the constraints on the CPU design. Since the CPU is only awakened for rare events, in many situations, it might not need to meet real-time requirements.
The present study remains a controlled system demonstration, not a clinical evaluation. Beat detection, segmentation, and FFT feature extraction are performed off-chip; the train/test split is beat-level, not patient-wise; and the energy analysis combines measured 130-nm front-end hardware with a projected 22-nm digital back end. These choices let us isolate the wake-up architecture and compare operating points on identical inputs, but future work should integrate the full sensing chain, use patient-independent evaluation, and validate a monolithic or heterogeneously integrated implementation.

Three directions are especially promising. (i) \emph{Smarter front ends:} larger probabilistic models, hierarchical cascades, and feature-extracting pre-stages that cut wake-ups without sacrificing sensitivity. (ii) \emph{Adaptive policies:} on-device adjustment of uncertainty thresholds and, where available, dynamic ($V_\mathrm{DD}$) to minimize measured wake-up frequency under task-level constraints; this can be framed as online bandit or reinforcement learning on a low-dimensional policy. (iii) \emph{Broader tasks and sensors:} audio, vibration, environmental monitoring, and vision with sparse events, where wake-up economics are most favorable. In all cases, the virtue of nanodevice Bayesian front ends is not software-equivalent accuracy per se, but trustworthy screening with native uncertainty that allows the programmable compute to sleep most of the time, precisely the behavior edge systems need for long autonomy.



\section*{Methods}

\subsection*{Design and fabrication of the memristor-based Bayesian machine}

The integrated circuit embeds 16 memristor arrays directly within the computing core of the logarithmic Bayesian machine. Its overall architecture is derived from the stochastic Bayesian processor of Ref.~\cite{harabi2023memristor}.

Each memristive memory array is organized in 64-bit words implemented in a two-transistor/two-memristor (2T2R) topology, as in Refs.~\cite{harabi2023memristor,jebali2024powering}. Access devices are thick-oxide transistors rated up to 5~V, which is required for both forming and programming of the memristors. The array periphery operates from three distinct power domains:
\begin{itemize}
    \item $V_\mathrm{DD}$ (nominally 1.2~V), which supplies the digital logic and sensing circuitry.
    \item $V_\mathrm{DDR}$ (up to 5~V), which biases the word lines connected to the gates of the selection transistors.
    \item $V_\mathrm{DDC}$ (up to 5~V), which biases the bit lines and source lines of the memory arrays.
\end{itemize}
Row- and column-level level shifters convert $V_\mathrm{DD}$-domain control signals to the higher $V_\mathrm{DDR}$ and $V_\mathrm{DDC}$ domains; these shifters are also implemented with thick-oxide devices. Readout is performed with precharge sense amplifiers (PCSAs) \cite{zhao2009high,zhao2014synchronous,harabi2023memristor}, designed using high-threshold thin-oxide transistors to minimize dynamic energy during operation. The layout of the memristor arrays and associated mixed-signal circuitry was drawn manually in Cadence Virtuoso and validated by circuit simulations using Siemens Eldo.

The digital blocks (decoders, adders, registers, and control logic) were described in SystemVerilog and synthesized with Cadence Genus using high-threshold thin-oxide transistors. A custom Tool Command Language (TCL) flow drives Cadence Innovus to perform fully automated placement and routing on top of a manually crafted floorplan. Standard physical-verification steps (design rule checking, layout-versus-schematic comparison, and antenna checks) were carried out with Calibre tools to ensure correct and reliable implementation.

The logarithmic Bayesian machine was fabricated using the same CMOS–memristor integration flow as in Refs.~\cite{harabi2023memristor,jebali2024powering,bonnet2023bringing}. The CMOS circuitry was produced by a commercial foundry in a low-power 130-nm technology featuring four metal layers. The memristive devices use a TiN/HfO$_x$/Ti/TiN material stack, where the HfO$_x$ layer is deposited by atomic layer deposition and the Ti layer is given a comparable thickness. The nominal device diameter is 300~nm. A fifth metal level is added on top of the arrays to contact the memristors, which are aligned above exposed vias in the underlying interconnect. Input/output pads are arranged along a single edge of the die and tailored to a custom probe card to facilitate electrical characterization.

\subsection*{Design of the wake-up platform with programmable RISC-V CPU}

The wake-up platform is described entirely in SystemVerilog. A single RTL code base is used both for experimental prototyping on FPGA and for the ASIC implementation used in the energy-evaluation study. In the experimental setup, the FPGA version of the platform is connected to the fabricated logarithmic Bayesian machine, which allows us to validate the wake-up flow and to study how device-and circuit-level behavior (in particular memristor non-idealities) translates to system-level performance.

The Bayesian machine is designed to operate in an always-on power domain and to wake up the rest of the system when required. The SoC comprises:
\begin{itemize}
    \item a CV32E40P CPU core ~\cite{schiavone2017slow}, which is a small, energy-efficient, 32-bit in-order RISC-V core with a 4-stage pipeline that implements the RV32IMFC instruction set (base integer I, integer multiplication and division M, compressed instructions C, and single-precision floating point F); 
    \item an AXI-Lite bus adapted from Ref.~\cite{alencar2025adam};
    \item general-purpose I/Os and a JTAG interface adapted from Ref.~\cite{alencar2025adam};
    \item a data memory (1~Mb) and a RISC-V program memory (1~Mb). On FPGA, these memories are implemented using on-chip Block RAM (BRAM) together with a behavioral memory model for simulation.
    \item a system configuration unit responsible for activity and clock gating of each SoC module; it also includes a programmable clock divider.
    \item a front-end controller for the Bayesian machine that can wake the back end and fetch Bayesian-machine inputs through a small DMA engine.

\end{itemize}

The RISC-V core accesses all subsystems (the Bayesian machine, data and program memories, and I/Os) through the AXI-Lite interconnect using memory-mapped addressing, which ensures that the same software stack runs unchanged on both the FPGA prototype and the ASIC target.

\subsection*{Memristor programming and characterization}

Before using the circuit for Bayesian inference, all memristive devices are first electroformed to create conductive filaments. During this forming step, the supply voltages are set to $V_\mathrm{DDC}$ = 3.5~V, $V_\mathrm{DDR}$ = 2.5~V, and $V_\mathrm{DD}$ = 1.2~V. Each device is then individually selected by the on-chip digital control logic, which applies a programming pulse of 0.5~\textmu s to the targeted memristor (see Supplementary Notes of Ref.~\cite{harabi2023memristor} for further details).

After forming, devices are programmed either to a low-resistance state (LRS) or a high-resistance state (HRS). In the SET operation (LRS programming), $V_\mathrm{DDC}$ is raised to 3.5~V while $V_\mathrm{DDR}$ is chosen at a specific compliance voltage. For RESET (HRS programming), $V_\mathrm{DDC}$ is increased to 3.25~V and $V_\mathrm{DDR}$ to 4.5~V. Programming pulses of opposite polarity are applied to each device depending on whether an LRS or HRS is targeted, following the same protocol as in Ref.~\cite{harabi2023memristor}.

In SET mode, the access transistor gate is biased at a compliance voltage ($V_\mathrm{DDR}$), which limits the compliance current and prevents overstress, thereby improving endurance and reliability. In RESET mode, current compliance is largely self-limited by the increasing device resistance as switching proceeds; in this case, the access transistor is driven up to 4.5~V at the gate, effectively turning it fully on and minimizing its influence on the RESET dynamics. 
On a separate chip optimized for fast programming experiments, we measured the statistics of the conductance of memristors programmed under these conditions for various compliance voltages; the resulting cumulative distributions are shown in Fig.~\ref{fig:machine}e.

The memory arrays use a two-transistor / two-memristor (2T2R) organization. Information is stored by complementary programming of the two devices: a logic ``0'' is encoded by programming the left memristor (connected to BL) to HRS and the right memristor (connected to BLb)  to LRS, whereas a logic ``1'' is encoded by programming the left device to LRS and the right device to HRS. This differential-like coding improves storage robustness against variability and noise \cite{hirtzlin2020digital}. The log-likelihood values required by the Bayesian inference engine are represented as 8-bit integers stored in this 2T2R array, as described in the main text.

\subsection*{Proof-of-concept heartbeat classification task}

We evaluate the wake-up system on a proof-of-concept heartbeat classification task derived from an electrocardiogram (ECG) database from MIT (MIT-BIH Arrhythmia Database) \cite{moody2000physionet}. 
We consider four classes: normal beats and three abnormal beat classes annotated L (left bundle branch block), R (right bundle branch block), and P (paced beats, denoted ``/'' in the original dataset).
For each annotated beat on both measuring channels, we extract a 700~ms segment centered on the QRS peak time annotation, and compute its frequency-domain representation by fast Fourier transform (FFT) with a 360~Hz sampling frequency.  Each extracted 700~ms segment constitutes one classification input to the wake-up system; because it is centered on one annotated heartbeat, we refer to it below simply as a beat.
The spectrum is then normalized by segment length, yielding spectral coefficients that are independent of window size.

We construct balanced training and test sets with an equal number of examples per class. The test set contains 800 beats per class; the training set contains 3,200 beats per class. The Bayesian front-end model and the multilayer perceptron (MLP) back-end model are trained on exactly the same training set.

\paragraph{Bayesian front-end model.}

The logarithmic Bayesian machine implements a discrete Bayesian classifier over a small set of spectral features. 
We compute $\chi^2$ scores on the training set (FFT bins from both ECG channels treated as distinct candidate features). We retain the four highest-ranking features for the hardware implementation.
For each retained bin, the magnitude is quantized into eight levels; the clipping ranges are chosen on the training set to maximize class separation. For each class and each quantized feature, we estimate the likelihood by histogramming the training data and smoothing the histogram with a Gaussian kernel. The resulting probabilities are then mapped to 8-bit log-domain values using
\begin{equation}
n=\mathrm{round}\!\left(m\log_B p\right), 
\end{equation}
so that $p \approx B^{n/m}$ in hardware. 
We use $B=0.15$ and $m=16$, which provided the best accuracy-quantization trade-off in our experiments. This mapping allocates seven code levels to probabilities above 0.5 and the remaining 249 to probabilities below 0.5, thereby providing finer resolution where the classifier most needs it.

\paragraph{MLP back-end model.}
The back-end model is a multilayer perceptron (MLP) that takes 32 quantized FFT features as input and has three fully connected layers of widths 74, 100, and 4. We again use $\chi^2$ ranking to select 32 frequency bins as inputs to the MLP. 
The network is trained in a quantization-aware manner, with a softmax output layer and a cross-entropy loss; at inference time we use the $\arg\max$ of the class probabilities as the predicted label.
The trained MLP is implemented in C on the RISC-V CPU using TensorFlow Lite with 8-bit signed integer quantization  \cite{david2021tensorflow}.  The program is compiled using the RISC-V GNU Toolchain (GCC, version 13.2.0) with an aggressive set of flags to optimize for code size and speed. Key compilation options include the architecture specification (\texttt{-march=rv32imfc\_zicsr}, \texttt{-mabi=ilp32f}), aggressive optimization (\texttt{-O3}), and memory-saving techniques such as enabling linker garbage collection (\texttt{-ffunction-sections}, \texttt{-fdata-sections}, and \texttt{-Wl,--gc-sections}) and forcing static memory allocation for TensorFlow (\texttt{-DTF\_LITE\_STATIC\_MEMORY}).

\paragraph{Wake-up policy.}
In the wake-up experiments, we stream the entire test set through the system beat-by-beat. For each beat, the Bayesian front end produces class log-scores and the wake-up controller applies this wake-up policy:
\begin{itemize}
    \item \emph{Wake-up on abnormal:} the RISC-V back end is awakened whenever the highest-scoring class of the Bayesian machine is one of the three abnormal classes.
    \item \emph{Wake-up on ambiguous/invalid:} the back end is also awakened when the Bayesian machine output is ambiguous or invalid. Specifically, this occurs when the highest-scoring class is ``normal'' but at least one abnormal class has the same score, or when one or more classes decode to zero output in the on-chip representation, which we interpret as a hardware or numerical failure. (Because the smoothed software model contains no exact zero probabilities, such zero hardware outputs are treated as invalid.)
\end{itemize}

\paragraph{Reset-based wake-up handling.}
In the implemented platform, wake-up is realized as a reset of the RISC-V CPU, not by invoking a conventional interrupt service routine. When the wake controller detects an abnormal or uncertainty condition, it asserts a wake/reset signal toward the CPU. Execution then restarts from the reset vector, and the first instructions of the firmware read a memory-mapped status register of the Bayesian accelerator/controller to determine whether the entry corresponds to a true startup or to a wake-up request. On a true startup, the firmware performs the normal platform initialization. On a wake-up request, it skips full initialization, reads the front-end status and class scores, executes the MLP inference, stores the final decision, and returns the system to low-power monitoring. In the ASIC implementation, the same wake/reset event also re-enables the previously clock- and power-gated digital domains before software execution begins.

For each heartbeat, the final system-level prediction is the label returned by the MLP when it is awakened, or the Bayesian machine prediction when no wake-up is triggered.

\subsection*{Details of the experimental wake-up proof-of-concept}

The logarithmic Bayesian machine is characterized under a semi-automatic Cascade PA200 probe station using a custom probe card that exposes all required power and I/O pads. The always-on inference supply of the Bayesian machine, $V_\mathrm{DD}$, is provided by a Keithley 2450 Sourcemeter and swept between 0.7~V and 1.2~V to emulate different energy-reliability operating points. 
During wake-up experiments, the domains $V_\mathrm{DDR}$ and $V_\mathrm{DDC}$ are not driven at programming voltages, but are tied to the regular front end supply voltage $V_\mathrm{DD}$, the only voltage swept during inference.
Throughout the manuscript, the label $V_\mathrm{DDR}$ in operating-condition sweeps therefore refers to the SET-compliance voltage used during the preceding memristor-programming step, not during inference.

All remaining components of the system (the RISC-V CPU, AXI-Lite interconnect, on-chip memories, and wake controller) are prototyped on a Xilinx zybo z7-20 FPGA board. The RTL of the SoC is synthesized with Vivado 2025 and mapped onto the board together with a thin hardware wrapper that drives the I/O pins connected to the probe card and ensures voltage conversion and ESD protection. 

The FPGA and the Bayesian machine communicate through a parallel memory-mapped interface that mirrors the AXI-Lite address map used in the ASIC design, ensuring the same software-visible register interface and RTL control flow as in the final SoC.
The FPGA prototype and the ASIC study use the same RTL and firmware but not the same clock frequencies. In the FPGA experiments, the Bayesian-machine interface was clocked at 2~MHz and the programmable back end at 50~MHz. These values were chosen pragmatically for the split experimental setup: 2~MHz shortens experimental runtime while remaining comfortably within the timing margin of the off-chip probe interface, whereas 50~MHz was the highest reliably validated back-end frequency on the FPGA platform. In the ASIC study, by contrast, post-layout simulations use 1~MHz for the always-on front-end side and 100~MHz for active back-end service. The FPGA prototype is therefore used for functional validation of the wake-up path, whereas absolute digital timing and energy projections are taken from the ASIC implementation.

A host computer runs a Python-based control and acquisition framework that programs the source-measure units over GPIB and records data from both the instruments and the FPGA over UART. To emulate a sensor, we implemented a wrapper that supplies the next prerecorded feature vector each time the system requests new input data.

For each ECG beat in the test dataset, the system:
\begin{enumerate}
    \item fetches the next input feature vector from the sensor-emulation wrapper and feeds it to the Bayesian machine for inference;
    \item runs front-end inference on that input vector with the Bayesian machine;
    \item applies the selected wake-up policy in hardware; if the policy does not trigger, the system finalizes the beat locally as normal;
    \item if the policy triggers, the wake controller asserts a reset-based wake-up signal to the RISC-V core and other components of the back end. The CPU restarts from the reset vector, the firmware distinguishes wake-up service from a true startup by reading a memory-mapped status register in the front end control logic, executes the MLP inference, records the final decision, and returns the system to low-power monitoring.
\end{enumerate}

On the FPGA prototype this sequence is reproduced with activity and clock gating, whereas in the ASIC design the same wake/reset event also restores the previously power- and clock-gated domains before the CPU begins executing.

All experimental runs use the same bitstream and firmware image; only the supply voltage $V_\mathrm{DD}$ of the Bayesian machine and the memristor programming conditions are varied between sweeps.

\subsection*{Energy evaluation of the complete wake-up system}

To quantify system-level energy beyond the split prototype, we combine measured front-end energy with post-layout estimates for the programmable digital domains. The digital platform, including the RISC-V core, SRAMs, AXI-Lite interconnect, wake controller, and power-management logic, was implemented in a 22~nm Fully Depleted Silicon-on-Insulator (FD-SOI) CMOS technology using the RTL described above. The memristor Bayesian machine, by contrast, was not re-designed in that node for this study; its contribution is derived from measurements of the fabricated 130-nm mixed-signal CMOS-memristor chip used in the experiments. The mean energy per input is then computed using eqs.~\eqref{eq:energie} and~\eqref{eq:monitorcost}.

The digital terms are obtained from post-layout gate-level simulations of the 22-nm implementation. We selected High Threshold Voltage (HVT) standard cells to optimize for leakage power. To extract realistic dynamic memory access costs during the physical implementation, we used Ultra-Low Power (ULP) SRAM macros as the available characterized memory option in the 22-nm PDK for both the 1~Mb instruction memory and the 1~Mb data memory. This choice is a modeling convenience, not a preferred deployment architecture. In a practical wake-up SoC, the instruction memory, which stores fixed firmware and neural-network weights, would more naturally be implemented in ROM or embedded non-volatile memory (NVM), allowing the block to be power-gated during idle monitoring without state loss. The data memory can remain SRAM, because it stores only transient execution variables and can also be safely power-gated between wake-up events. Since the dynamic read energy of embedded NVMs is comparable in magnitude to that of SRAM \cite{inci2021deepnvm++}, we used SRAM dynamic access costs as a proxy for active inference for both memories, while excluding memory static power from system-level idle estimates to reflect the intended power-gating strategy.

The SystemVerilog RTL, identical to that used in the FPGA prototype, was synthesized using Synopsys Design Compiler. Physical implementation, including floorplanning, clock tree synthesis, and placement and routing (PnR), was performed using Cadence Innovus.

To capture realistic power profiles, we performed post-layout gate-level simulations using Siemens ModelSim. These simulations were delay-annotated using Standard Delay Format (SDF) information extracted from the physical layout. The testbench executed the complete heartbeat classification firmware, generating switching activity files (Value Change Dump) that capture the precise transitions of the logic and memory during idle monitoring, wake-up-induced restoration of the clock- and power-gated domains, reset-vector execution, status-register checking, and active neural network inference.

Final power consumption was analyzed using Synopsys PrimePower, which combined the extracted parasitics with the simulated switching activity (Fig.~\ref{fig:asic}d), to obtain the energy consumption of a wake-up service event $E_\mathrm{service}$ and the power consumption during monitoring standby $P_\mathrm{mon}$.
 For the voltage-scaling analysis, we use the actual measurements on the fabricated Bayesian machine (Fig.~\ref{fig:asic}b) for the energy consumption of the Bayesian machine. 
 We  assume that the power consumption  $P_\mathrm{mon}$  of the front end support logic scales with front end supply voltage $V_\mathrm{DD}$.  Starting from the nominal post-layout value of $2.9~\mu\mathrm{W}$, with 55\% static power and 45\% dynamic power, we model the static contribution as scaling linearly with $V_\mathrm{DD}$ and the dynamic contribution as scaling quadratically with $V_\mathrm{DD}$. The corresponding monitoring-energy term is obtained by multiplying the resulting $P_\mathrm{mon}$ by the monitoring period $T_s$: $E_\mathrm{mon} = P_\mathrm{mon}T_s$.


\section*{Acknowledgements}
This work  benefited from  France 2030 government grants managed by the French National Research Agency (ANR-22-PEEL-0013, ANR-22-PEEL-0010, ANR-23-PEIA-0009, ANR-24-RRII-0004).  The authors would like to thank L.~Hutin and J.~Grollier for discussion and invaluable feedback.   A large language model (OpenAI ChatGPT) was used for copyediting parts of this manuscript.

\section*{Author contributions statement}
T.B. designed and performed all the experiments. The digital platform was initially designed by A.R. and F.P.A. under the guidance of P.B. and D.N. It was adapted to nanodevices by T.B. T.B. and A.R. performed the ASIC energy analysis. B.L.F., T.D., and T.B. wrote the RISC-V C code. A.R. and M.A.I. performed device-level characterization and optimization. C.T. and K.E.H. designed the logarithmic Bayesian machine under the guidance of J.M.P. and D.Q.  C.T. designed the printed circuit board.  E.V.  led the fabrication of the logarithmic Bayesian machine. D.N. and D.Q. supervised the work. D.Q. wrote the initial version of the manuscript. All authors discussed the results and reviewed the manuscript. 

\section*{Competing interests}
The authors declare no competing interests.

\section*{Data availability}
The MIT-BIH Arrhythmia Database analyzed in this study is publicly available \cite{moody2000physionet}. 
The data measured in this study are available from the corresponding author upon request.

\section*{Code availability} 
The RTL code of the CV32E40P CPU core is publicly available \cite{schiavone2017slow}.
The RTL code of the AXI-Lite interconnect and JTAG modules are publicly available \cite{alencar2025adam}.
The C code for MLP inference on the RISC-V CPU is available from the corresponding author upon request.
The software implementations of the machine-learning models used in this work are available from the corresponding author upon request.

\bibliography{references}  

\FloatBarrier

\end{document}